\documentclass[10pt,leqno]{amsart}
\usepackage{graphicx}
\usepackage{indentfirst,csquotes}
\usepackage{subcaption}
\usepackage{tikz}
\usepackage{pgfplots}
\usepackage{multirow}
\usepackage{array}     
\usepackage{longtable}   
\usepackage{booktabs}    
\usepackage{float}
\usepackage{placeins}
\usepgfplotslibrary{groupplots} 

\usepackage{amssymb,amsthm,amsmath}
\usepackage{xcolor,paralist,fancyhdr,etoolbox}

\begin{document}
\title{Dense or Sparse? \\ Post-Packing Interconnection Analysis in FPGAs}
\author{Xiaoke Wang and Dirk Stroobandt}
\maketitle
\begin{center}
  {\small $^1$Ghent University, Belgium} \\
  {\small \texttt{xiaoke.wang@ugent.be}, \texttt{dirk.stroobandt@ugent.be}}
\end{center}
\date{\today}

\let\thefootnote\relax
% \footnotetext{MSC2020: Primary 00A05, Secondary 00A66.} %%%%%%%%%%

\begin{abstract}
Packing is a crucial step of FPGA design, directly impacting interconnect complexity, routing congestion, and overall performance. This paper presents a post-packing interconnect-aware analysis, illustrating how dense (sparse) packing changes the interconnection structure. We introduce a new metric, RDensity, to define post-packing density and investigate its influence on routability. Through a comparative study of two packing tools, we demonstrate that density has a direct impact on routability. Our findings provide valuable insights into how packing decisions affect FPGA efficiency and offer guidance for improving FPGA packing tools and architecture design by integrating interconnect-aware methods. The goal is to achieve efficient routing while maintaining an optimal balance between cluster density, CLB pin counts, and logical block sizes.
\end{abstract}
\section{Introduction}

The FPGA physical design flow consists of several key stages, including \emph{packing}, which groups primitives such as Look-Up Tables (LUTs) and Flip-Flops (FFs) into larger Configurable Logic Blocks (CLBs) before placement and routing~\cite{goos_vpr_1997, betz_architecture_1999}. Packing directly impacts routing congestion, wirelength, and timing closure, making it a crucial optimization step in FPGA Computer-Aided Design (CAD). Traditional packing strategies primarily focus on maximizing CLB utilization. Building upon the conventional understanding, \emph{Dense packing} minimizes the number of clusters by aggressively filling logic blocks, which can increase local connectivity but often leads to severe routing congestion~\cite{murray_vtr_2020, luu_vtr_2014}. Conversely, \emph{sparse packing} can spread logic more evenly, improving routability at the cost of increased placement complexity. The trade-off between packing density and routability remains a fundamental challenge in FPGA CAD.

To address these challenges, two primary packing methodologies have been developed: \emph{seed-based packing} and \emph{partition-based packing}. Seed-based methods, such as VPack~\cite{betz_architecture_1999}, iteratively build clusters around an initial seed primitive based on connectivity heuristics. Extensions like T-VPack~\cite{marquardt1999using} and AAPack~\cite{Luu2011AAPack} incorporate timing awareness and routability constraints~\cite{chen_efficient_2014}. However, these methods often result in overly dense clusters, exacerbating routing difficulties. Partition-based approaches, including PPack~\cite{feng2012kway, feng_rents_2014}, pre-assign logic elements to partitions before clustering, improving routability through hierarchical bipartitioning~\cite{vercruyce_how_2018}. Although effective, these partition-based methods lack adaptability across different FPGA architectures and heterogeneous embedded blocks, which is the reason seed-based is broadly used. Breaking Boundaries enables re-clustering for better timing and wirelength~\cite{elgammal_breaking_2023}, while VIPER corrects infeasible placements to reduce re-packing~\cite{thurmer_viper_2024}. These approaches emphasize the need for post-packing optimization to enhance placement and routing.

\begin{figure}[ht]
    \centering
    \includegraphics[width=.9\linewidth]{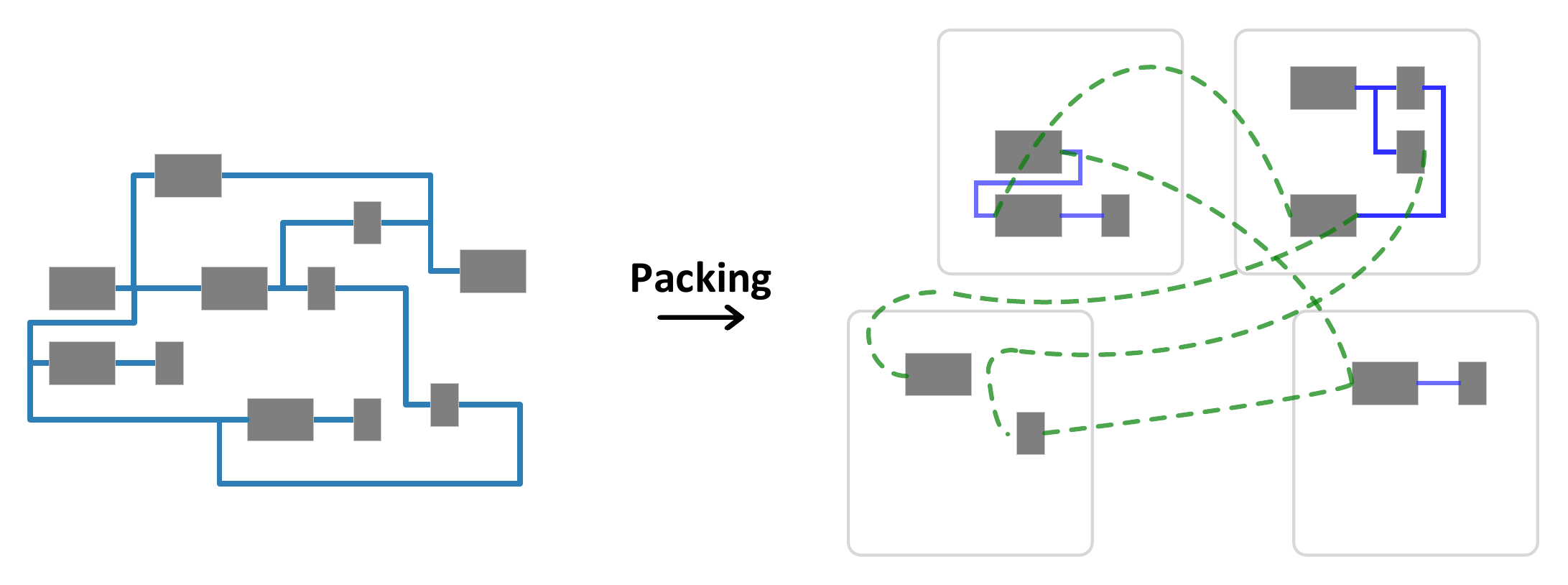}
    \caption{Packing process: pre-packing netlist (left) and post-packing netlist (right).}
    \label{fig:packing}
\end{figure}

Despite these efforts, a critical gap remains in understanding how packing affects FPGA placement and routing at the interconnection level. Previous work~\cite{wang2025infl} demonstrated that interconnection complexity has a double-exponential impact on routing runtime and a single-exponential impact on placement. Packing modifies interconnection topology by aggregating logic elements into clusters, reorganizing interconnections into two levels: \emph{inter-CLB} and \emph{intra-CLB}. As illustrated in Fig.~\ref{fig:packing}, inter-CLB connections (green) span multiple CLBs, while intra-CLB connections (purple) exist within individual CLBs. The FPGA interconnection architecture imposes constraints on terminal availability, particularly for interleaved pins at the CLB level, requiring packing algorithms to restructure interconnections accordingly.

This study provides an analysis of post-packing interconnect structures and their influence on routing. We propose a new definition of dense (sparse) packing, and a metric to describe density. We evaluate the difference between the two packing tools, focusing on their density's impact on wirelength, and congestion.  The remainder of this paper is structured as follows. Section~\ref{bg} reviews the analysis methods of interconnection structures and refers to the application of the current post-packing scenario. Section~\ref{Methodology} describes our experimental methodology and offers examples. Section~\ref{Results&Discussion} compares the packing results of two methods and analyzes the interconnection changes and their influence. Section~\ref{Conclusion&fw} presents the results and discussion and concludes the findings. It also outlines future research plans.

\section{Background and Theoretical Framework}
\label{bg}

\subsection{Rent's Rule}

Rent’s rule~\cite{landman1971pin, christie2000rent} describes an empirical relationship between the number of external connections (\emph{terminals}) and the number of internal logic elements within a circuit partition. It is widely used to characterize interconnect complexity in digital circuits. The relationship is expressed as:
\begin{equation}
    T = t \cdot B^r
\label{eq:1}
\end{equation}
where \( T \) represents the number of external terminals of a partition, \( B \) is the number of internal logic elements, \( t \) is the average number of terminals per block, and \( r \) is the Rent exponent. The exponent \( r \) indicates the interconnection complexity, with values close to 1 implying a highly interconnected network, whereas lower values suggest more sparse connectivity.

% In this study, we leverage Rent’s rule to analyze the \textbf{post-packing net} and compare the observed Rent exponent for \textbf{dense} and \textbf{sparse} packing strategies. By computing the \emph{partition-observed} Rent exponent, we aim to quantify the interconnect differences introduced by these packing methodologies. 

% A higher Rent exponent after packing suggests increased interconnect complexity and potential routing challenges, while a lower exponent indicates more localized connectivity and improved routability. This metric provides valuable insight into how packing affects the downstream FPGA placement and routing stages.

\subsection{Post-Packing Interconnection Complexity}
To analyze the interconnection structure, we proposed two separate levels: inter-CLB and intra-CLB. This distinction also applies to other blocks at the same level as CLBs, such as DSP, RAM, and IO blocks. As the theory model of the interconnection structure is the same for all other different blocks, we use only soft logic benchmarks and analyze dense and sparse packing only on CLBs\footnotemark. \footnotetext{CLB mentioned in this paper is also called logic array block (LAB) in Intel Stratix architecture.}

\subsubsection{Definition of Packing Density}
Based on conventional wisdom, packing density quantifies the utilization of resources within Configurable Logic Blocks (CLBs). A higher packing density often improves resource efficiency but can lead to routability challenges. It is formally defined as:
\[
D_X = \frac{\overline{X}_{clb}}{X^{tot}_{clb}}, \quad X \in \{B, T\}
\]
where \( B \) represents the number of logic blocks used in CLBs, and \( T \) represents the number of terminals. The numerator corresponds to the average used resources (\(\overline{B}/\overline{T}\)), while the denominator refers to the total available blocks (for \( B \)) or pins (for \( T \)) in each CLB.

\subsubsection{Definition of Rent Packing Density (RDensity)}
 Routability is highly dependent on the circuit's interconnection structure. Therefore, the classification of dense or sparse packing should not be solely based on utilization but must also account for the number of interconnections. Using the Rent's rule, we can illustrate this concept as follows. We first estimate the average size (i.e., the number of basic logic blocks used) of the CLBs based on the pre-packing interconnection complexity from Equation~\ref{eq:1}:
\[
\overline{B}_{clb}^* = \left(\frac{\overline{T}_{clb}}{t}\right)^{\frac{1}{r}}
\]
where $\overline{T}_{clb}$ is the average number of terminals used in CLBs. Comparing this estimated average size with the actual average size of the CLBs ($\overline{B}_{clb}$), we define:
\begin{itemize}
    \item \textbf{RDense Packing:} \( \overline{B}_{clb} > \overline{B}_{clb}^* \)
    \item \textbf{RSparse Packing:} \( \overline{B}_{clb} < \overline{B}_{clb}^* \)
    \item  \textbf{RModerate Packing:} \( \overline{B}_{clb} \approx \overline{B}_{clb}^* \)
\end{itemize}
To quantitatively assess whether a packing strategy results in excessive interconnection density or sparsity, we define the \textit{Rent Packing Density} (RDensity) parameter:
\begin{equation}
D_{R}= \frac{\overline{B}_{clb}}{\overline{B}_{clb}^*}
\label{eq:packing_density}
\end{equation}

\begin{figure}[ht]
    \centering
    \includegraphics[width=.8\linewidth]{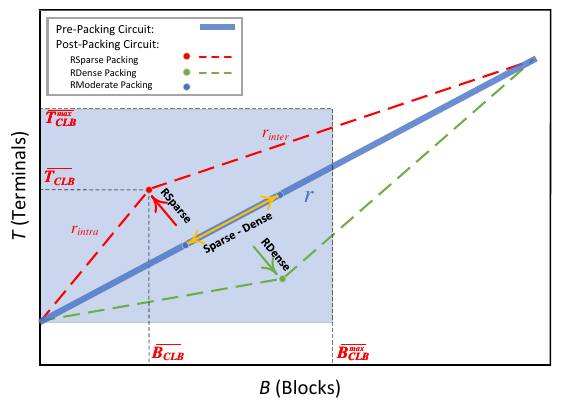}
    \caption{Rent's Rule diagram illustrating RDense and RSparse packing; A demonstration of interconnect complexity.}
    \label{fig:DorL}
\end{figure}
As shown in Fig.~\ref{fig:DorL}, in the logarithmic scale, based on the equation~\ref{eq:1}, the relationship between the terminals and the size of the partitions becomes linear. However, the packing process will break this uniform relationship and transform the original Rent's curve (blue) into two distinct segments with different slopes, corresponding to intra- and inter-CLB interconnections because of the architectural differences. The “inflection point”—the location where the slope changes (located in the blue region)—serves as an indicator of packing density: when it shifts above the pre-packing curve, the packing is RSparse (i.e, the number of blocks used inside the CLB is smaller than expected from the number of terminals used); when it shifts below, the packing is RDense (i.e, the number of blocks used inside the CLB is larger than expected from the number of terminals used). In contrast, if the inflection point remains on the original curve, the deviation—whether trending upward or downward (yellow arrow)—does not reflect a true change in RDensity of packing. We may characterize it as exhibiting sparse or dense packing by using more or fewer intra-CLB resources but with the same RDensity. An ideal packing algorithm should achieve a moderate packing RDensity, maximizing resource utilization within CLBs (dense). This balance ensures efficient use of FPGA resources without unnecessarily increasing inter- and intra-CLB interconnection complexity.

\begin{figure*}[ht]
  \centering
  \begin{subfigure}{0.3\textwidth}
    \centering
    \includegraphics[width=\linewidth]{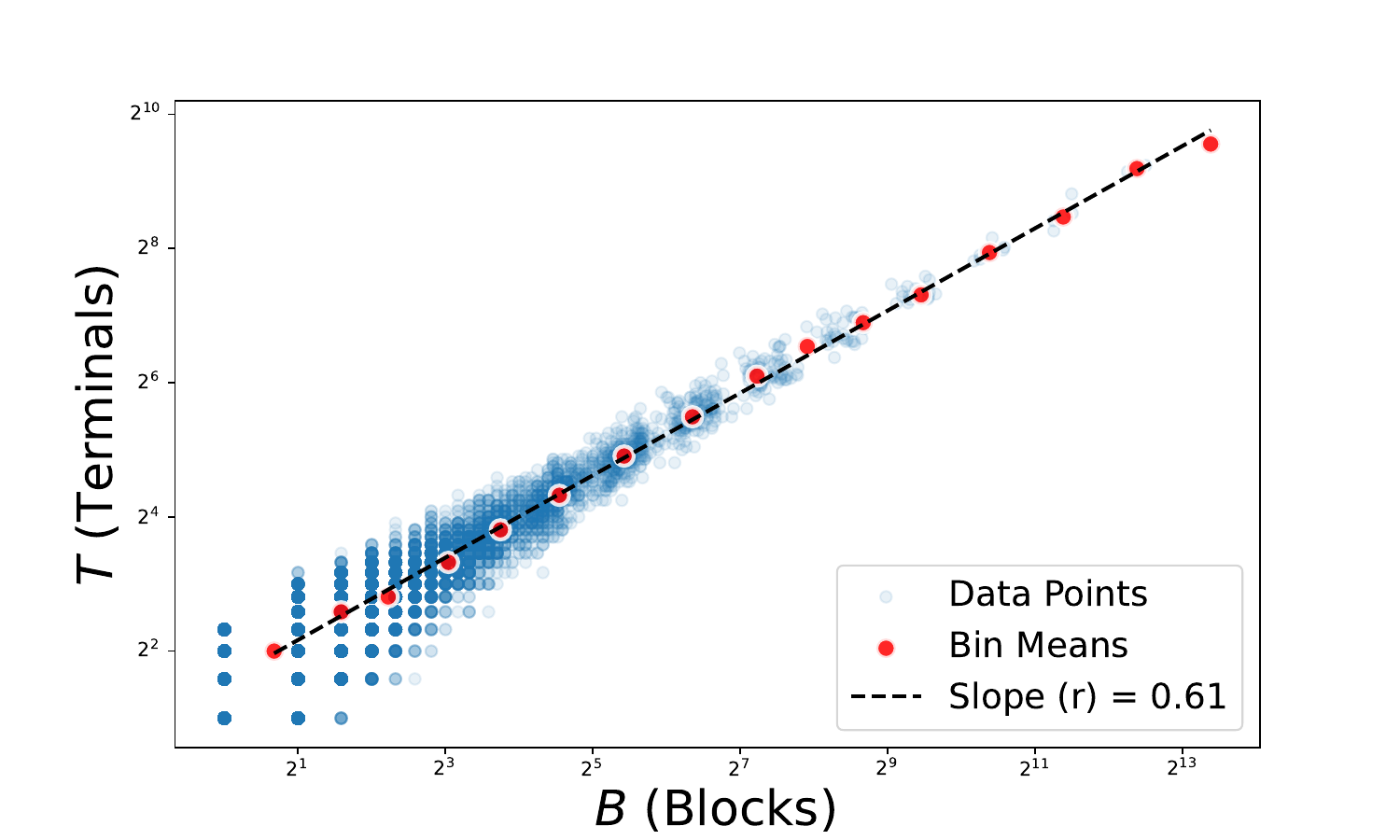}
    \caption{Netlist (\emph{.blif}).}
    \label{r_(a)}
  \end{subfigure}
  \begin{subfigure}{0.3\textwidth}
    \centering
    \includegraphics[width=\linewidth]{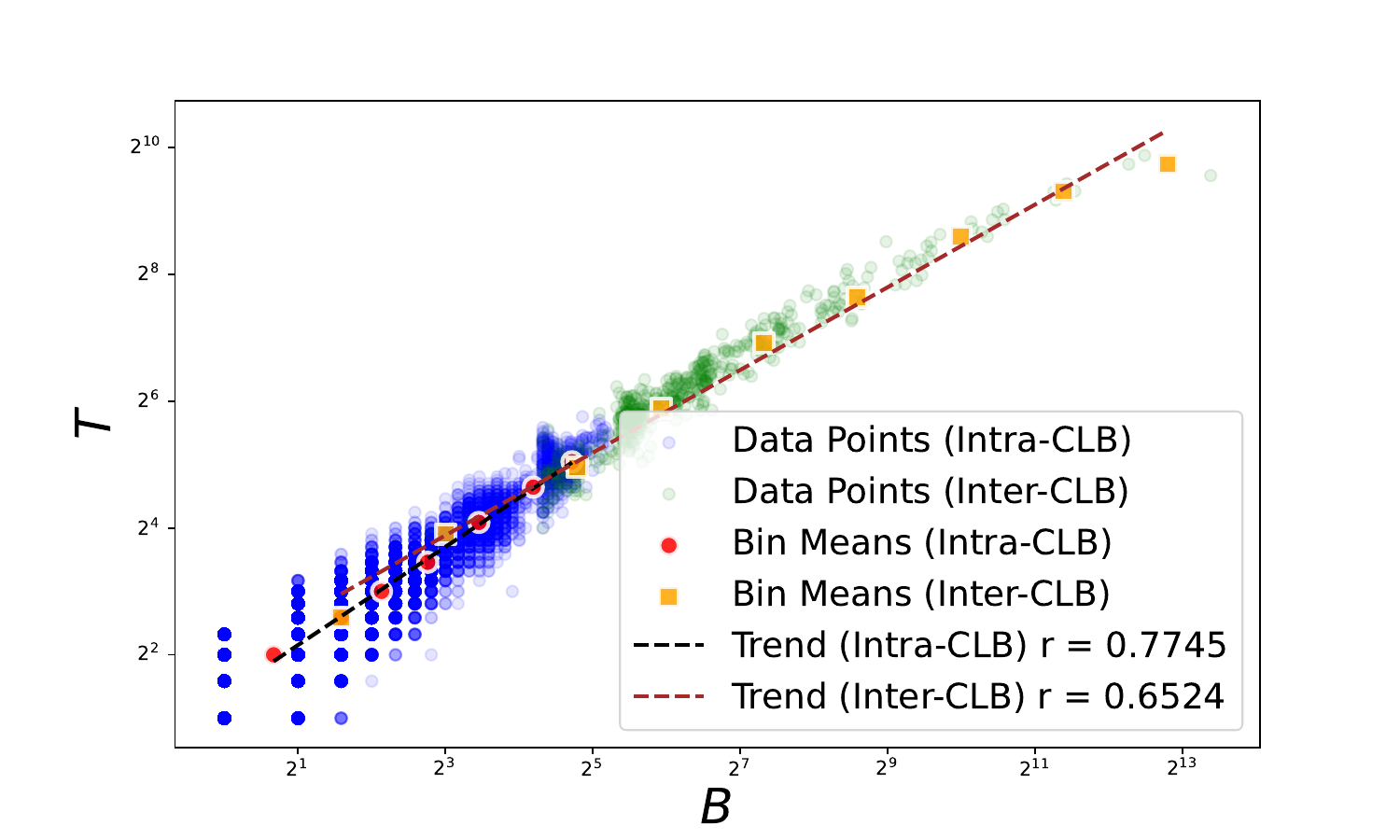}
    \caption{Dense RSparse.}
    \label{r_(b)}
  \end{subfigure}
  \begin{subfigure}{0.3\textwidth}
    \centering
    \includegraphics[width=\linewidth]{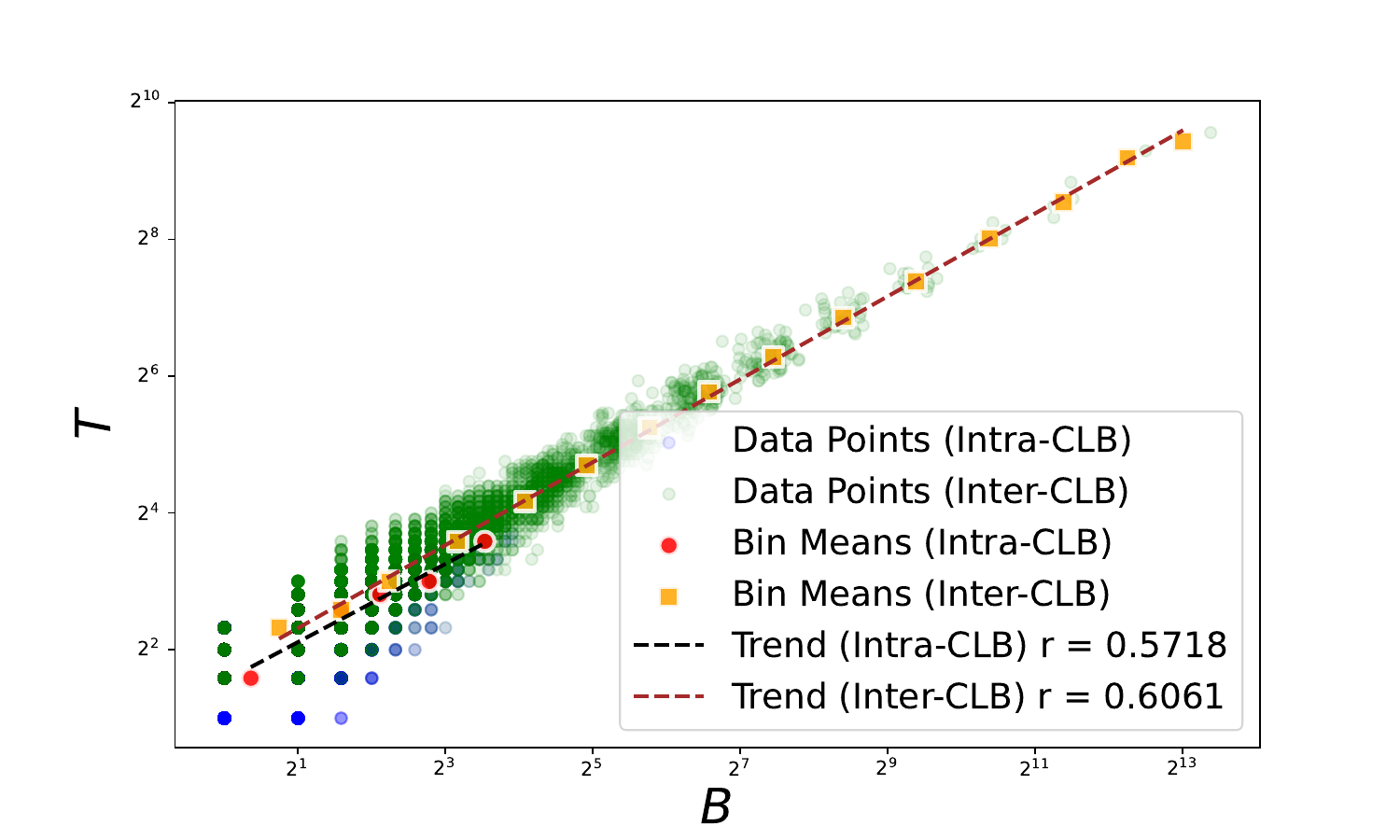}
    \caption{Sparse RSparse.}
    \label{r_(c)}
  \end{subfigure}
  \caption{Visualization of Rent's rule (pre-packing and post-packing).}
  % \Description{Detailed description of images for accessibility.}
  \label{fig:inter_intra_rent}
\end{figure*}

\section{Methodology}
\label{Methodology}
\subsection{Rent Plot}
Using the VTR 8 project, by going through the AAPack algorithm, we obtained the post-packing netlist in \emph{.net} format~\cite{murray_vtr_2020}. After reconstructing the connections as a graph, we applied \texttt{hMetis} \cite{karypis1998hypergraph}, a hypergraph partitioning package, to perform recursive bi-partitioning. This process allowed us to generate a Rent’s rule plot, the same as the Fig.~\ref{fig:DorL}, showing the relationship between terminals and block sizes, from which the Rent exponent can be easily observed. 
\subsection{Post-Packing Analysis}
We start with a netlist that has an original Rent exponent of $0.6$ and perform a post-packing analysis. Fig.~\ref{r_(a)} presents the original Rent plot. For the post-packing circuit, we separately plot the relationship between terminals and blocks for inter- and intra-CLB partitions, as shown in Figs.~\ref{r_(b)} and \ref{r_(c)}. These results are obtained using AAPack on the same circuit with two extreme target block pin utilization constraints ($1.0$ and $0.0$ for \texttt{target\_ext\_pin\_util}). By comparing the outcomes, we can observe the differences between the two configurations.
The intra-CLB and inter-CLB regions exhibit linear regressions with different slopes, corresponding to two distinct Rent exponents. The packer mostly changed the density, and only slightly changed the RDensity.

For these RSparse packing results, the packer tries to limit the number of terminals and therefore cannot add more blocks than the interconnection complexity allows, resulting in an RSparse density. Consequently, the interleave region (inflection point) between the two levels is slightly above the previous linear regression line, as shown in Figure~\ref{r_(b)}, leading to a significantly higher intra-CLB slope compared to earlier.
The RSparse packing raises the starting (inflection) point of the inter-CLB curve in Rent’s diagram. If the endpoint remains fixed, this should lead to a lower slope, suggesting that the circuit is simpler outside the CLBs. However, measurements contradict this hypothesis. Regardless of the inflection point’s location, the inter-CLB curve maintains the same slope or even increases slightly and forms a distinct second region on the right side as shown in Fig.~\ref{r_(b)}, ultimately returning to the original number of terminals. The inter-CLB slope remains largely unchanged, as it is primarily determined by the original interconnection complexity.

The Fig.~\ref{r_(c)} in a relative RDense and sparse packing, then this packer configuration can pack on average a little more blocks into the CLBs than expected, resulting in a (slightly) large $D_R$. We see then that the intra-Rent exponent is somewhat lower, but the inter-Rent remains almost the same.

\section{Results and Discussion}
\label{Results&Discussion}

To evaluate the impact of packing strategies, we compare the Rent packing density parameter $D_R$ between VTR 8 and VTR 7. The primary differences between these two versions lie in their handling of unrelated clustering, connection-driven clustering, and packing prioritization criteria. These factors significantly influence both packing density and then routability. We use benchmarks generated by GNL \cite{stroobandt1999generating, stroobandt1999towards, verplaetse2000synthetic} with controlled complexity and some MCNC benchmarks \cite{yang1991logic}. The comparison results are presented in Table~\ref{tab:vpr_comparison}, tracking total wirelength ($TWL$), routing runtime ($RT_{route}$), CLB usage, average CLB terminal utilization ($\overline{T}_{clb}$), inter- and intra-CLB Rent exponents ($r_{inter}$ and $r_{intra}$) observed using the weighted method~\cite{wang2025infl}, and the Rent density factor ($D_R$).
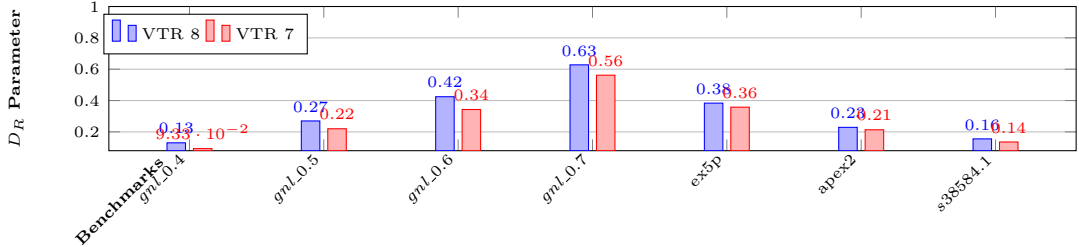
\begin{figure*}[htbp]
\centering
\begin{tikzpicture}
\begin{axis}[
    width=\linewidth, 
    height=3.5cm,
    bar width=7pt,
    ylabel={\textbf{$D_R$ Parameter}},
    xlabel={\textbf{Benchmarks}},
    xlabel style={at={(0,0)}, rotate=45, font=\tiny},
    symbolic x coords={$gnl\_0.4$, $gnl\_0.5$, $gnl\_0.6$, $gnl\_0.7$, ex5p, apex2, $s38584.1$},
    xtick=data,
    x tick label style={rotate=45, anchor=east, font=\tiny},
    ymajorgrids=true,
    legend style={at={(0.1,0.95)}, anchor=north,legend columns=2, font=\tiny}, 
    enlarge x limits=0.1,
    ymin=0.08, ymax=1.0,
    nodes near coords,
    font=\tiny,
    ybar, 
    bar shift=0pt, 
    xtick align=inside,
    group style={
        group name=my bar plot,
        group size=2 by 1, 
        x descriptions at=edge bottom,
        horizontal sep=1pt 
    }
]
\addplot+[
    bar shift=-5pt, 
    fill=blue!30
] coordinates {($gnl\_0.4$,0.1302)
($gnl\_0.5$,0.2698)
($gnl\_0.6$,0.4244)
($gnl\_0.7$,0.6278)
(ex5p, 0.3835)
(apex2, 0.2290)
($s38584.1$, 0.1554)};

\addplot+[
    bar shift=5pt,
    fill=red!30
] coordinates {($gnl\_0.4$,0.0933)
($gnl\_0.5$,0.2200)
($gnl\_0.6$,0.3428)
($gnl\_0.7$,0.5618)
(ex5p,0.3578)
(apex2,0.2134)
($s38584.1$,0.1356)};

\legend{VTR 8, VTR 7}
\end{axis}
\end{tikzpicture}
\vspace{-5mm}
\caption{Comparison of the $D_R$ parameter between VTR 8 (blue) and VTR 7 (red) across different benchmarks.}
\label{fig:bar_P}
\end{figure*}

\begin{table*}[htbp]
\centering
\caption{Comparison of VTR 8 and VTR 7 Across Different Benchmarks}
\label{tab:vpr_comparison}
\renewcommand{\arraystretch}{0.8}
\small
\begin{tabular}{|c|c|c|c|c|c|c|c|c|c|c|}
\hline
\textbf{Benchmark} & \textbf{$r_{netlist}$} & \textbf{Tool} & \textbf{$TWL$} & \textbf{$RT_{route}$ (s)} & \textbf{CLB Usage} & \textbf{$\overline{T}_{clb}$} & \textbf{$r_{inter}$} & \textbf{$r_{intra}$} & \textbf{$D_R$} \\
\hline
\multirow{2}{*}{gnl\_0.4} & \multirow{2}{*}{0.4} & VTR 8 & 31870 & 0.49 & 509 & 22.68 & 0.4231 & 0.5837 & 0.1302 \\
                          &                        & VTR 7 & 41454 & 0.61 & 450 & 27.23 & 0.5588 & 0.6241 & 0.0933 \\
\hline
\multirow{2}{*}{gnl\_0.5} & \multirow{2}{*}{0.5} & VTR 8 & 45376 & 0.63 & 487 & 26.83 & 0.5511 & 0.6345 & 0.2698 \\
                          &                         & VTR 7 & 57334 & 0.76 & 450 & 30.91 & 0.6452 & 0.6705 & 0.2200 \\
\hline
\multirow{2}{*}{gnl\_0.6} & \multirow{2}{*}{0.6} & VTR 8 & 104032 & 0.97 & 476 & 32.40 & 0.6575 & 0.6981 & 0.4244 \\
                          &                         & VTR 7 & 116830 & 1.20 & 450 & 36.28 & 0.7102 & 0.6994 & 0.3428 \\
\hline
\multirow{2}{*}{gnl\_0.7} & \multirow{2}{*}{0.7} & VTR 8 & 407064 & 2.33 & 485 & 36.38 & 0.7951 & 0.7531 & 0.6278 \\
                          &                         & VTR 7 & 423650 & 2.83 & 451 & 41.37 & 0.8356 & 0.7888 & 0.5618 \\
\hline
\multirow{2}{*}{ex5p} & \multirow{2}{*}{0.67} &  VTR 8 & 8256 & 0.23 & 64 & 16.62 & 0.5763 & 0.7123 & 0.3835 \\
                               &                        & VTR 7 & 8305 & 0.23 & 58 & 18.34 & 0.5773 & 0.7214 & 0.3578 \\ 
\hline
\multirow{2}{*}{apex2} & \multirow{2}{*}{0.59} &  VTR 8 & 14761 & 0.28 & 105 & 39.25 & 0.4805 & 0.7350 & 0.2290 \\
                               &                        & VTR 7 & 16522 & 0.29 & 104 & 41.14 & 0.5448 & 0.7403 & 0.2134 \\ 
\hline
\multirow{2}{*}{s38584.1} & \multirow{2}{*}{0.48} & VTR 8 & 38083 & 0.80 & 329 & 22.60 & 0.3924 & 0.5910 & 0.1554 \\
                          &                        & VTR 7 & 56266 & 1.51 & 310 & 23.99 & 0.7323 & 0.6108 & 0.1356 \\
\hline
\end{tabular}
\end{table*}

\subsection{Impact of Packing RDensity}
As shown in Fig.~\ref{fig:bar_P}, the RDensity factor $D_R$ for VTR 8 is consistently greater than that of VTR 7 and is closer to $1$, indicating a more balanced packing approach. However, all results suggest a \textbf{RSparse} packing configuration, as they are significantly lower than $1$. This indicates that pins are a bottleneck while many internal resources remain unused. Further investigation is needed to determine whether this is due to the architectural constraints within a CLB or a limitation of the packer itself.
The observed differences in CLB usage and terminal count between VTR 8 and VTR 7 can be attributed to their distinct packing strategies. VTR 8, by disabling unrelated clustering and enforcing connection-driven clustering, produces fewer but more functionally coherent clusters. In contrast, VTR 7 packed unrelated clusters into the same CLB, resulting in a lower number of CLBs. However, VTR 8's packing ensures that clusters maintain stronger internal connectivity, thereby reducing excessive inter-cluster connections. Even VTR 8 uses more CLBs and each has less size, but it's still relatively RDenser than VTR 7 as the terminals and block size collaborate well. 

The results from Table~\ref{tab:vpr_comparison} indicate that VTR 8 has better results than VTR 7. It is an established and well-documented fact that VTR 8 demonstrates superior performance over VTR 7. The advantages and differences between these versions have already been analyzed in detail \cite{murray_vtr_2020}. 
However, by employing our approach, we aim to demonstrate that the factor $D_R$ we introduce is an effective and reliable method for quantifying and evaluating the RDensity of packing algorithms. Even though VTR 8 demonstrates improvements, it still deviates from an RModerate packing strategy with a small $D_R$, a deviation that becomes even more prominent in circuits with lower interconnection complexity. This underscores the significance of the proposed metric in enhancing FPGA packing strategies.

% Additionally, VTR 8 prioritizes transitive connectivity and employs a balanced cluster seed selection strategy, which helps distribute logic elements more effectively. As a result, while VTR 8 exhibits a lower CLB usage and fewer terminals per cluster, the $D_R$ parameter remains lower than VTR 7, indicating that the packing is less sparse despite using fewer blocks.

% The higher density in VTR 8 does not translate to excessive routing congestion, as the reduced inter-cluster connectivity compensates for the smaller cluster count. This suggests that the balance between packing density and routability in VTR 8 is better optimized, leading to more efficient use of logic and interconnect resources compared to VTR 7.

\subsection{Impact of Inter- (Intra-) Rent Exponent}

The results indicate that VTR 8 optimizes both inter-CLB and intra-CLB Rent exponents simultaneously. This suggests that the compensation for intra-CLB interconnections in RSparser packing (VTR 7) does not translate into any benefits for inter-cluster connectivity. Instead, it leads to the emergence of a pronounced second region, which can significantly increase the inter-CLB Rent exponent. Consequently, a high \( r_{\text{inter}} \) negatively impacts routability, increasing routing congestion and overall FPGA inefficiency.

\subsection{Insights for Architecture}
The inflection point is constrained by the maximum available terminals and logic resources within CLBs, defining a theoretical boundary that encompasses the moderately strong packing scenario. For example, in Fig.~\ref{fig:DorL}, the top-right (maximum) corner of the blue region is typically above most circuit curves in modern FPGA architectures. This indicates that the FPGA architecture consistently encourages the packer to perform RSparse packing, which partially explains why all results exhibit RSparse characteristics. Consequently, co-designing FPGA architectures alongside packing algorithms is crucial to achieving an optimal balance between logic utilization and interconnect efficiency.

% \subsection{Conclusion}
% The results demonstrate that while VTR 7 achieves the highest packing density, it does so at the cost of increased routing complexity. VTR 8, by implementing connection-driven clustering and prioritizing transitive connectivity, achieves a $D_R$ closer to 1, indicating a more balanced trade-off between packing efficiency and routability. These findings highlight the importance of structured clustering in FPGA CAD tools, suggesting that logical locality and interconnect optimization

\section{Conclusion and Future Work}
\label{Conclusion&fw}

% The routing time follows a double-exponential dependency on inter-CLB complexity and a linear dependency on circuit size (number of CLBs) \cite{wang2025infl}. Sparse packing improves interconnect complexity but increases block usage, leading to a scaling effect. This creates a trade-off between minimizing inter-CLB complexity and controlling the number of CLBs, both of which significantly impact overall design efficiency.

In this work, we introduced a new dense/sparse theory together with a novel metric for evaluating the RDensity of FPGA packing algorithms by integrating both the number of terminals and block size. This metric provides a more comprehensive assessment of packing efficiency than traditional methods, which often focus solely on utilization or logical density. 
Through our comparative study of VTR 8 and VTR 7, we demonstrated that a well-balanced packing algorithm must carefully manage both cluster size and external pin count to achieve optimal density while maintaining routability. This suggests that emphasizing connection-driven packing and maintaining a good post-packing interconnection structure is essential for achieving both high-utilization and high-performance FPGA implementations. The insights from our metric have practical implications for both FPGA CAD tool development and FPGA architecture design.

Some more measurements are needed to evaluate the density of the packing results. Combining the intra- and inter-CLB interconnection complexity with density $D_R$, we can well illustrate and evaluate how packing will change the interconnection structure and influence the routability.
Based on those findings, we will further explore the packing algorithm by incorporating additional methods to fine-tune the interconnection structure between inter- and intra-CLB levels. Rather than relying solely on AAPack, we will analyze and compare other packing approaches to assess their effectiveness in optimizing the post-packing interconnection structure. Our goal is to develop a more interconnection-aware packing tool that can better adapt to modern FPGA architectures.



\bibliographystyle{ACM-Reference-Format}
\bibliography{heart-bib}


\end{document}